\documentclass[12pt,preprint]{aastex}
\usepackage{graphicx}
\usepackage{subfigure}
\usepackage{mathrsfs,amssymb}
\usepackage{amsmath}
\usepackage{natbib}
\usepackage{color}
\usepackage{ulem}

\newcommand{\Tv}{\ensuremath{T_{\rm vir}}}

\newcommand{\K}{\ensuremath{{\rm K}}}

\newcommand{\keV}{\ensuremath{\, {\rm keV}}}

\newcommand{\Msun}{\ensuremath{{M_\sun}}}

\begin{document}

\title{Reionization in the Warm Dark Matter Model}

\author{Bin Yue\altaffilmark{1,2}, 
Xuelei Chen \altaffilmark{1,3}}
\altaffiltext{1}{National Astronomical Observatories, Chinese Academy of Sciences, 
20A Datun Road, Chaoyang, Beijing 100012, China}
\altaffiltext{2}{Graduate University of Chinese Academy of Sciences, Beijing 100049,
China}
\altaffiltext{3}{Center of High Energy Physics, Peking University, Beijing 100871, China}

\begin{abstract}
Compared with the cold dark matter (CDM) model, in the warm dark matter (WDM) model formation 
of small scale structure is suppressed. It is often thought that this would delay the 
reionization of the intergalactic medium (IGM), as the star formation rate during the 
epoch of reionization (EOR) would be lowered. However, during the later 
stage of the EOR, a large portion of the ionizing photons are consumed by recombination 
inside the minihalos, where the gas has higher density and 
recombination rates than the gas in the IGM. The suppression of small scale 
structure would 
therefore reduce the recombination rate, and could potentially 
shorten the reionization process. This effect is investigated here by using the analytical
``bubble model'' of reionization. We find that in some cases, though the initiation of the 
EOR is delayed in the WDM model, its completion could be even earlier than the CDM case,
but the effect is generally small. 
We obtain limits on the WDM particles mass for different reionization redshifts.
\end{abstract}


\maketitle

\section {Introduction}\label{intro}
The nature of the dark matter is presently unknown. 
The cold dark matter (CDM) model has been very successful in explaining the 
observed properties of large scale structures of the Universe, but on small scales
there are still some discrepancies, as the abundance of satellite galaxies falls far short
of the number of subhalos predicted 
(\citealt{1999ApJ...522...82K,1999ApJ...524L..19M,2008MNRAS.391.1685S};
but see also \citealt{2007ApJ...670..313S} and \citealt{2009NJPh...11j5029P}
for a different view). 
An interesting alternative is the warm dark matter (WDM) scenario, where the dark matter 
particle has a smaller (a few keV) mass 
\citep{1982Natur.299...37B,1982PhRvL..48..223P,1985ApJ...298....1M,1986ApJ...304...15B,1994PhRvL..72...17D,1996ApJ...458....1C,
2000ApJ...542..622C,2001ApJ...551..608S}. As the growth of structure below the free-streaming
scale of the WDM was quenched, such a model predicts much fewer small halos, 
but on large scales its prediction is similar to the CDM case and 
could still match the large scale structure observations very well 
\citep{2001ApJ...556...93B,2001MPLA...16.1795J,2006PhRvD..73f3513A,2008MNRAS.386.1029K,2009MNRAS.399.1611T,2010MNRAS.404L..16M,2011PhRvD..84f3507S,2011PhRvD..83d3506P}. 

The abundance of small halos may also greatly affect the reionization of the Universe. 
As the stars and galaxies form at the end of the cosmic 
dark age, the star light ionizes the 
intergalactic medium (IGM), causing the later to ionize. Eventually, all of 
hydrogen gas in the IGM is ionized, this is the so called (hydrogen)
reionization \citep{2001PhR...349..125B,2006ARA&A..44..415F}. In the 
hierarchical structure formation scenario, most of these ionizing photons would come
from dwarf galaxies in small halos. 
In recent years, the Gunn-Peterson trough in the Lyman alpha absorption spectrum
of high redshift quasars has been observed, indicating that we are approaching 
the end stage of the epoch of reionization (EOR) at $z>6$
\citep{2001AJ....122.2850B,2002AJ....123.1247F}. The cosmic microwave background (CMB)
anisotropy observation is another important source of information about 
EOR \citep{1997PhRvD..55.1822Z,2003ApJS..148..161K,2007ApJS..170..377S}. The 
recent WMAP analysis gives the redshift of completion of reionization as 
 $z_{re}=10.5 \pm 1.2$ \citep{2011ApJS..192...16L}. This result can be 
used to constrain the 
properties of dark matters. In the WDM model, halos of a given mass would typically 
form later than in the CDM model, so the initiation of the 
EOR would be delayed. Also, as the formation 
of small halos which are most common during the EOR are suppressed, fewer
ionizing photons are produced, and it was believed that this would 
delay the completion of the reionization 
significantly \citep{2001ApJ...558..482B,2003ApJ...591L...1Y,2003ApJ...593..616S}. 

It should be noted that in some WDM models, the WDM particle may be able to decay,
which ionizes and heats up the IGM \citep{2001ApJ...562..593A,2005MNRAS.364....2M,
2006MNRAS.369.1719M}. This (and the annihilation of some CDM particles)
in itself is an interesting problem with rich physics involved
(see e.g. 
\citealt{2004PhRvD..70d3502C,2006PhRvD..74j3502F,2008ApJ...679L..65C,2010JCAP...10..023Y,
2010MNRAS.406.2605R}). The effect on reionization is fairly complicated, 
for example, the free electrons produced in the dark age could help catalyze the formation of 
molecule hydrogen, making it possible to form first stars and start reionization 
early in this case \citep{2006PhRvL..96i1301B,2007NuPhS.173...24K,2007ApJ...654..290S}, 
though the magnitude and significance of this effect is highly 
uncertain\citep{2007MNRAS.374.1067R,2007MNRAS.375.1399R}.
In the present paper, we shall limit ourselves to the passive WDM model, and shall not consider
such effects.

However, the time it took the reionization process to complete depends not only on 
the supply rate but also on the consumption rate of the ionizing photons. 
After being ionized, the atoms of the IGM gas will recombine, and to keep the
gas ionized more ionizing photons have to be consumed. As the 
recombination rate is proportional
to the density squared, at the later stage of the EOR 
the minihalos could potentially consume the majority of the ionizing 
photons \citep{2001ApJ...551..599H,2001MNRAS.320..153B,2002ApJ...578....1B,2004MNRAS.348..753S,
2004ogci.conf..549I,2005MNRAS.361..405I,
2006MNRAS.366..689C,2009MNRAS.398.2122Y,2010arXiv1003.6132A}.
In the WDM scenario, the number of minihalos are much fewer, so the
global recombination rate could be significantly lower. When this is considered, it is not
obvious whether the WDM would delay or advance the completion of the EOR. 

We use the bubble model of reionization \citep{2004ApJ...613....1F} to 
investigate the reionization process in the WDM scenario, take into
account both the reduction in photon production rate and the consumption rate due to the 
suppression on halo formation in WDM models.
Here we adopt the WMAP 5 years cosmology parameters (the WMAP 7 years 
parameters are almost identical):
$(\Omega_m, \Omega_{\Lambda}, \Omega_b, h, \sigma_8, n_s)
= (0.274, 0.726, 0.0456, 0.705, 0.812, 0.95)$ \citep{2009ApJS..180..330K}. Note that 
for the WDM model, this set of parameters may not be the best fit values, 
but to illustrate the physical effect of different dark matter mass, 
we have used the same parameters for all models.

\section{Method of Calculation}\label{methods}

\begin{figure}[htbp]
\begin{center}
\includegraphics[scale=0.4]{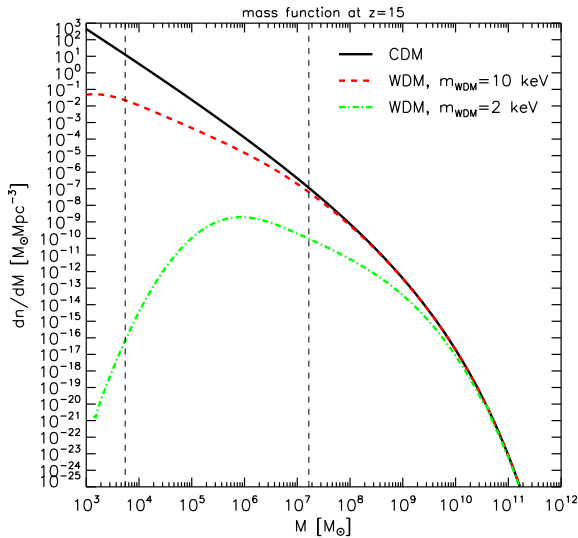} 
\caption{The halo masses function for CDM (solid line) and 
WDM with $m_{\rm WDM}$=10, 2~keV (dashed, dashed-dotted) respectively at z=15.
We marked the minihalos mass range by two thin dashed vertical lines.}
\label{mf}
\end{center}
\end{figure}

We calculate the halo mass function in the WDM model analytically 
by following the prescription of 
\citet{2011PhRvD..84f3507S}. In the WDM model, 
the free-streaming comoving scale is given by
\begin{equation}
\lambda_{fs}\approx 0.11\left(\frac{\Omega_{\rm WDM}h^2}{0.15}\right)^{1/3}
\left(\frac{m_{\rm WDM}}{\rm keV}\right)^{-4/3}\rm Mpc,
\end{equation}
and the corresponding mass is 
$M_{fs}=4\pi/3(\lambda_{fs}/2)^3\rho_m$. 
The halo mass function is then given by
\begin{equation}
\frac{dn}{dM}(M,z)=\frac{1}{2}\left\{ 1+{\rm erf}\left[\frac{{\rm log_{10}}(M/M_{fs})}
{\sigma_{\rm logM}}\right] \right\}
\left[\frac{dn}{dM}\right]_{PS},
\label{dndm2}
\end{equation}
where $\sigma_{\rm logM}=0.5$, and $\left[\frac{dn}{dM}\right]_{PS}$ can be calculated 
with the usual Press-Schechter (PS)
prescription \citep{1974ApJ...187..425P,1991ApJ...379..440B}, with the 
matter power spectrum for the WDM case given by the 
fit \citep{2001ApJ...556...93B,2005PhRvD..71f3534V}:
\begin{equation}
P_{\rm WDM}(k)=P_{\rm CDM}(k)\{[1+(\alpha k)^{2\mu}]^{-5/\mu}\}^2,
\label{pwdm}
\end{equation}
where $\mu=1.12$, and 
\begin{equation}
\alpha=0.049
\left(\frac{m_{\rm WDM}}{\rm keV}\right)^{-1.11}\left(\frac{\Omega_{\rm WDM}}{0.25}\right)^{0.15}
\left(\frac{h}{0.7}\right)^{1.22}h^{-1} \rm Mpc.
\end{equation}
The halo mass functions for the CDM model and the WDM model 
with different particle masses 
at redshift 15 are plotted in Fig. \ref{mf}. We see the abundance of dark 
matter halos on small scales are significantly suppressed, while on large scales
the halo abundance are almost the same as the CDM model. On the same plot, we also 
mark the mass range of what we call the ``minihalos'', i.e. halos which are sufficiently 
massive to accrete the gas, but not massive enough for the gas to cool by atomic H and form 
galaxies\citep{2006MNRAS.366..689C}.

We now consider the reionization process in the WDM model. 
At the end of the dark age, dark matter halos form from primordial overdense
perturbations, and in halos of sufficient masses, the 
gas cooled radiatively and contracted to dense clumps which 
eventually form the first stars. 
At present, it is still unclear whether the first stars formed in halos which 
were cooled by the trace amount of molecule hydrogen in the gas  
($\Tv >\sim10^3~\K$), or only in more massive halos which were cooled by atomic hydrogen 
($\Tv > 10^4~\K$) \citep{2001PhR...349..125B,2011arXiv1102.4638B}. 
In either case,  the stars and galaxies tend to form more abundantly 
in overdense regions, so the ionized regions first appeared as ``bubbles'' in such regions.
As more and more bubbles appeared and grew in size, eventually the ionized regions 
overlapped, thus completing the reionization.

Inspired by numerical simulation results, a popular analytical model of the process 
called the ``bubble model'' has been 
developed \citep{2004ApJ...608..622Z,2004ApJ...613....1F, 2004ApJ...613...16F}.
According to this model, the reionization proceeds by first forming 
ionized bubbles, the number and size of these bubbles continue to grow till
they overlap with each other, thus complete the reionization process.
The formation of the ionized bubbles is determined by the condition that the 
number of ionizing photons produced within the given region 
exceeds the total number of atoms to be ionized
in the same region. If we assume that each collapsed baryon on 
average contribute $\zeta$ ionizing photons, with each photon ionizes one atom once,
and the average number of recombination per atom during that time is $n_{rec}$, 
then the condition for the region to be ionized at the given redshift 
can be written as 
\begin{equation}
\zeta f_{coll}>1 +n_{rec}, 
\label{eq:f_recomb}
\end{equation}
where $f_{coll}$ is the fraction of baryons collapsed into 
star-forming halos. This treatment of the effect of 
recombination differs from \citet{2005MNRAS.363.1031F}. 
They considered the limiting case, where the ionizing 
photon production is required to counteract the recombination 
at that instant. Our model in Eq.~(\ref{eq:f_recomb}) is more in line
with the original bubble model, in which the total 
integrated number of ionizing photons is considered.

In such a model, one assumes that  halos with the virial masses 
above a certain threshold value could form stars.
The collapse fraction can then be calculated with the 
extended Press-Schechter 
method \citep{1991ApJ...379..440B,1993MNRAS.262..627L,1996MNRAS.282..347M}.
During the early EOR, the Pop III stars formed in the molecule 
hydrogen  cooled halos may 
have played an important role,  these halos are much less massive than the 
typical star-forming halos during the EOR, which are cooled by atomic hydrogen
or at somewhat later time, by metals. To account for this, we may split
the contribution into the part from molecule hydrogen cooled halos and the part
from more massive halos. We denote
the first type by the subscript ``mol'', and without danger of confusion, 
the second type is denoted without any subscript, then the condition Eq.(\ref{eq:f_recomb})
can be rewritten as 
\begin{equation}
\zeta_{mol}f_{mol}+\zeta f > 1+n_{rec}.
\label{Nrec}
\end{equation}

A molecule hydrogen cooled halo should have a virial temperature of at least
$10^3~\K$ for this mechanism to work. Furthermore, the formation of the molecules
is strongly modulated by the Lyman-Werner (LW) radiation background. For the 
molecules to form, the mass of the halo should be greater 
than \citep{2009ApJ...694..879T} 
\begin{equation}
M_{\rm H_2,cool}(J_{21},z)=6.44\times10^6J_{21}^{0.457}(\frac{1+z}{31})^{-3.557}~\Msun.
\end{equation}
The mass threshold is then given by 
\begin{equation}
M_{mol}(z)={\rm max}[M_{vir}(10^3~\K,z),M_{\rm H_2,cool}(J_{21},z)].
\end{equation}
We model the LW intensity as $J=J_{21}\times10^{-21}\rm ergs^{-1}cm^{-2}Hz^{-1}sr^{-1}$
with 
\begin{equation}
J_{21}(z)=10^{-3}\left(\frac{\zeta_{mol}}{50}\right)+0.28\left(\frac{\zeta}{40}\right)(1+z)^3f_{coll}(z), 
\end{equation}
The first term represents a slowly-evolving component due to Pop III stars whose
formation is self-regulated, while the second term comes from the more
massive, atomically cooled halos. Strictly speaking, for the WDM model the formation of 
Pop III stars is suppressed, so the coefficient of the first term would be smaller, 
but this change would not affect the final result, as the contribution of the few 
stars in WDM model is very small anyway.
The molecule cooled halo collapse fraction is then calculated with 
\begin{equation}
f_{mol}=\frac{1}{\rho_m}\int dz^\prime\int_{M_{mol}(z^\prime)}^{M_{vir}(10^4~K,z^\prime)}M\frac{d^2n}{dMdz^\prime}dM.
\end{equation}

The value of $\zeta$ and $\zeta_{mol}$ depends on a number of 
factors, e.g. the fraction of baryons in the halo which ended up in 
stars, the energy released by the star during its lifetime, and the 
fraction of the photons which escaped to the IGM from the cloud 
surrounding the star. There are currently large 
uncertainties on this parameter. Here we adopt $\zeta=40$ \citep{2001PhR...349..125B} and
$\zeta_{mol}=50$, 
the latter parameter value is in agreement with the properties of massive metal-free stars 
given in \cite{2002A&A...382...28S}. With these parameter values, $J_{21}\approx 10^{-3}$ 
during the Pop III stars dominated stage in the CDM model, in well agreement
with the value widely adopted in analytical and numerical studies of first stars \citep{2007ApJ...671.1559W}, 
while at z=10 $J_{21}\approx 7$, very close to 
the value given in \cite{2009ApJ...694..879T}.

At high redshifts, the clustering in the IGM is relatively low, 
we assume that only one photon is needed to ionize an IGM atom,
while for the atoms in the minihalo 
the number of recombinations is given by   
\begin{equation}
n_{rec,MH}=\frac{1}{\rho_m}\int_{M_J(z)}^{M_{up}(z)}\xi M\frac{dn}{dM}dM. 
\label{nrecMHs}
\end{equation}
Here $\xi$ is the average number of recombinations per atom of the minihalo,
$M_J$ is the Jeans mass. 
The upper limit of the integration $M_{up}$  
is the minimum mass of halos that could host radiation sources, i.e., Pop III 
stars or galaxies, that is 
${\rm min}[M_{mol}(z),M_{vir}(10^4~K,z)]$.
The recombination inside the galaxies should be relegated to 
the net photon production number $\zeta$ and should not be included again here.
Earlier analytical estimates typically gave $\xi \sim 10^2$ \citep{2001ApJ...551..599H},
in which case the recombination of the minihalos would consume the majority of photons 
at the late stage of EOR. However, 
numerical simulation of the ionizing front passing through minihalos shows that the 
number of actual recombinations may be far smaller, e.g., at z=15, 
$\xi < 8$ for halos with mass below
$10^7~\Msun$ and irradiated by typical ionizing flux
\citep{2004MNRAS.348..753S,2005MNRAS.361..405I}. As the halo is ionized, 
it is heated and the gas expands, causing the density to decrease and recombination rate
quickly lowered, and thus the recombination rate is much lower than the earlier estimates. 
We adopt the $\xi$ value \footnote{Our definition of $\xi$ differs by one from 
\cite{2005MNRAS.361..405I}: $\xi=\xi_{\rm Iliev}-1$, where $\xi_{\rm Iliev}$ is the value
given in \cite{2005MNRAS.361..405I}.} 
derived from \cite{2004MNRAS.348..753S} and \cite{2005MNRAS.361..405I}.

In Eq. (\ref{Nrec}), for a bubble with mass $m$, 
$f$ and $f_{mol}$ are both functions of the linear overdensity of this bubble. 
By solving the equation $\zeta_{mol}f_{mol}(z,\delta_x)+\zeta f(z,\delta_x)=1+n_{rec}$, 
a critical overdensity $\delta_x(m,z)$ is determined.
This is the barrier in the excursion set, regions with linear overdensity above this 
barrier should be ionized \citep{2004ApJ...613....1F}. 
We obtained this barrier by solving the above equation
with numerical iterations, and found that it is still well approximated by 
a linear function of the squared variance of density fluctuation $\sigma^2(m)$
(the following process is the same as in \citealt{2004ApJ...613....1F}):
$\delta_x(m,z)\approx B(m,z)=B_0+B_1\sigma^2(m)$, where $B_0=\delta_x(m\rightarrow\infty,z)$
and $B_1=\frac{\partial \delta_x}{\partial \sigma^2}(m\rightarrow\infty,z).$
With this linear barrier, the bubble mass function is expressed analytically 
by 
\begin{equation}
\frac{dn_{b}}{dm}=\sqrt{\frac{2}{\pi}}\frac{\rho_m}{m^2}\left |{\frac{d{\rm ln}\sigma}{d{\rm ln}m}}\right|
\frac{B_0}{\sigma(m)}{\rm exp}\left[-\frac{B^2(m,z)}{2\sigma^2{m}} \right],
\end{equation}
where $n_b$ is the number density of bubbles. Finally, the volume filling 
factor of all bubbles $Q_V$ is calculated directly by integrating over the bubble
mass function:
\begin{equation}
Q_V=\int V(m)\frac{dn_b}{dm}dm.
\end{equation}

\section{Results and Discussions}\label{results}

\begin{figure}[htb]
\begin{center}
\includegraphics[width=0.48\textwidth]{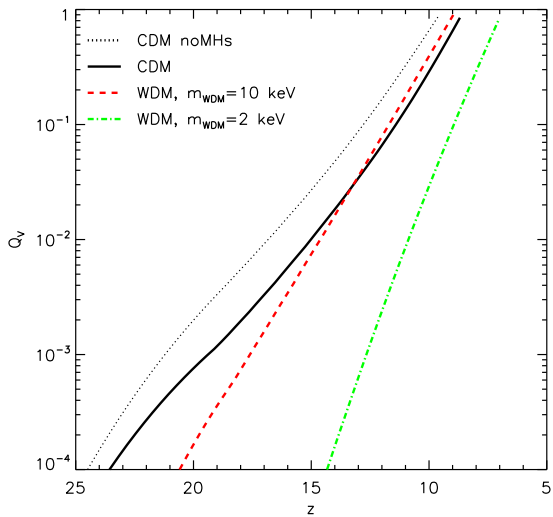}
\end{center}
\caption{The redshift evolution of the ionized bubble volume filling 
factor in different models.}
\label{QV_Iliev}
\end{figure}

We plot the redshift evolution of the ionization volume filling factor in 
Fig.~\ref{QV_Iliev}. Here we considered four different cases: the
reionization in the $\Lambda$CDM model with and without minihalo recombinations,  
and the reionization in $\Lambda$WDM model with minihalo recombinations  
for $m_{\rm WDM}=10~\keV$ and $2~\keV$.

First we note that as illustrated in the $\Lambda$CDM case, 
in the absence of minihalos, the reionization would be completed earlier
by as much as $\Delta z=1$, this shows how much impact the minihalo
recombinations could have on the reionization. This is obtained with $\xi$
derived from \citet{2005MNRAS.361..405I}. If we adopt the 
higher values of $\xi$ given in the earlier literature \citep{2001ApJ...551..599H},
the impact would be even stronger.

In the $\Lambda$WDM case, the reionization starts later than 
in the $\Lambda$CDM cases, as the first stars would form
later with the small scale power suppressed in these models. 
This difference is most obvious
at $z>20$, when the first stars just begin to form in large numbers. 
As small scale powers are suppressed in the 
$\Lambda$WDM model, there are very few halos in which the gas could be cooled by molecule 
hydrogen and form Pop III stars \citep{2006ApJ...648...31O}. Instead, in $\Lambda$WDM 
most first stars only form in atomically cooled halos,
this caused significant difference in the initiation of reionization. 

However, it is generally believed 
that the reionization is due primarily not to the stars formed in 
the H$_2$ cooled halos, but to stars formed in the more massive halos 
at lower redshifts. 
For the more massive halos, the difference between 
the $10 \keV$ WDM and the CDM is not so large.
In fact, from the figure we can see that in this case, the 
bubble filling factor of the $10\keV$ WDM model catches up with 
the CDM model at $z \sim 13$, and later it even exceeds that of the 
CDM model, thus 
the reionization is actually completed earlier than
in the CDM case. This is because at this stage, the ionizing photons are
produced mainly in the more massive halos, which is about equally abundant in 
the 10 keV WDM and CDM cases, while the suppression of minihalos  
reduced the global recombination rate in the WDM model, 
making the bubbles overlap earlier. Thus we observe the interesting result that 
the Universe is reionized earlier in the WDM model. 
This effect is relatively small compared with the 
current theoretical and observational uncertainties, especially the mean number of 
ionizing photons produced by a halo of given mass. However, it could potentially be useful
if these uncertainties are greatly reduced by improvements on observations and theoretical
modelings of reionization process.

For models with still lower WDM masses, the more massive halos are begin to be 
affected. In the 2~keV WDM model, the star formation associated with the more
massive halos are also suppressed, and the reionization is delayed 
in that model. The reduction of photon supply and consumption would 
be balanced somewhere between
these two cases, the exact value would however 
be dependent on the parameters we adopt, especially the value of 
the parameters $\zeta$ and $\xi$.
The values we adopted here are plausible and non-extreme, with the reionization 
happen at $z \sim 10$, in agreement with the WMAP constraint. 
Nevertheless, there are large uncertainties in these parameters. In particular, 
if the $\xi$ value for the minihalos is greater, as was used in some earlier 
papers, the effect would be still stronger.

\begin{figure}[ht]
\begin{center}
\includegraphics[width=0.48\textwidth]{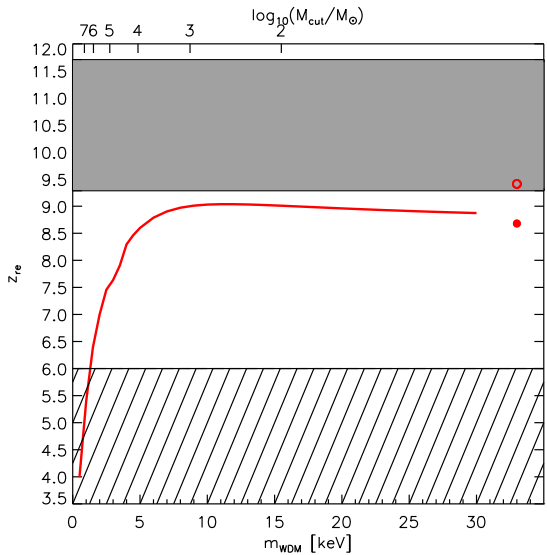}
\end{center}
\caption{The redshift of the completion of reionization 
as a function of $m_{\rm WDM}$ (the solid line).
On the right side of the figure, the open circle represents the result for the CDM model 
without minihalos, while the filled circle represents
the result for the CDM model when the effect of minihalo recombination is taken into
account. For comparison, the mass scale $M_{\rm cut}$, where the mass function 
$dn/dM$ is $1/e$ of the LCDM value, is also marked on the upper abscissa
of the figure. 
The shadowed region corresponds to $z_{re} < 6.0$, 
already excluded by the observations of Gunn-Peterson trough in $z\sim 6$ quasar absorption line
spectra. The gray region above is the 
constraint from the WMAP, i.e. $z_{re}=10.5\pm1.2$. Here we adopted the fiducial value $\zeta=40$.}
\label{zre}
\end{figure}

We plot the redshift of completion of reionization as a function of WDM
particle mass in Fig. \ref{zre}. For reference, the corresponding cutoff mass scale $M_{\rm cut}$,
where the mass function $dn/dM$ is suppressed by a factor of $e$
when compared with the $\Lambda$CDM model, is also plotted on the upper abscissa.
We see that below $11 \keV$, the reionization redshift is 
very sensitive to the WDM particle mass. With $m_{\rm WDM}$ increases, the reionization redshift
also raises quickly, as was usually assumed for WDM particles.  
However, a peak of reionization redshift of $z_{re}\approx9 $ is reached at
$m_{\rm WDM}\approx 11\keV$. Above this mass, the reionization redshift begin to decrease 
slowly, as the suppression on number of star-forming halos becomes relatively insignificant, 
while the suppression of minihalos reduced the global recombination rate.
As a comparison, we also plot $z_{re}$ in the CDM model with and without minihalo 
recombinations by the filled and open circles respectively. Quasar absorption line 
studies show that 
the Universe had been reionized at least $\sim6$ \citep{2006ARA&A..44..415F}.
This gives the constraint that $m_{\rm WDM} > 1.3$~keV. 

This limit on WDM mass may be compared with recent WDM mass limits obtained from 
other observations. For example, \citet{2000ApJ...543L.103N} obtained $m_{\rm WDM} > 0.75$~kev
from the Ly$\alpha$ forest observations, while \citet{2005PhRvD..71f3534V} obtained a lower 
limit of 0.55 kev from CMB (WMAP) and the Ly$\alpha$ forest data.
\citet{2001ApJ...558..482B} gave $m_{\rm WDM} > 1.2 \keV$ 
 with the requirement $z_{re} > 5.8$ for their fiducial model, while
\citet{2011PhRvD..83d3506P} obtained a constraint of $m_{\rm WDM} > 2.3 \keV$ 
from the number of Milky Way satellites.

\begin{figure}[htb]
\begin{center}
\includegraphics[width=0.48\textwidth]{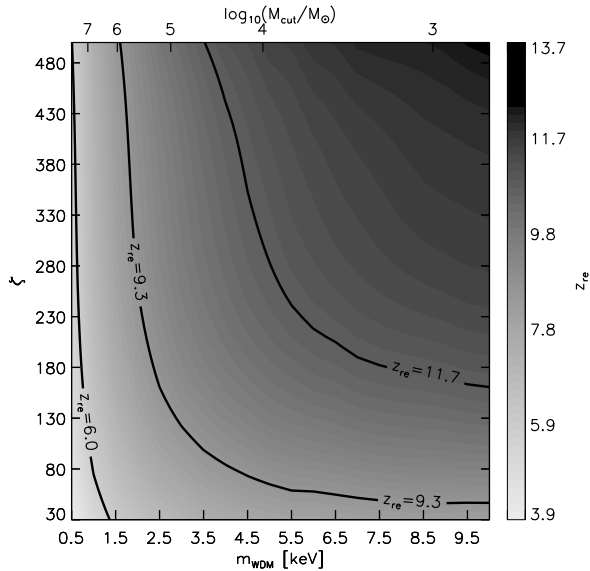}
\end{center}
\caption{The contour map of the redshift when reionization is completed in the 
$m_{\rm WDM}$-$\zeta$ plane, we label three lines, $z_{re}=6$, which is the lower limit of reionization
redshift given by
Gunn-Peterson trough observations, and $z_{re}=10.5\pm1.2$, which are from WMAP.}
\label{zre2}
\end{figure}

In the above, we have adopt the same value of $\zeta$ as used in the fiducial 
model of \cite{2001ApJ...558..482B}, which was derived from the observations of 
$z\sim3-4$ and present-day galaxies. However, there are still large uncertainties
on the properties of sources in the epoch of reionization. 
If these sources are stars with metallicity $Z=5\times10^{-4}~Z_\odot$,
and if their initial mass function is the Salpeter form \citep{1955ApJ...121..161S}
with the mass range $1 < M <100\Msun$, for the starburst model the number of ionizing photons 
produced per stellar atom is $\approx 13000$ \citep{2003A&A...397..527S}.
Assuming a star formation efficiency of 0.05 and escape fraction 0.5, $\zeta$ could be 
as high as $\approx 300$. Considering these uncertainties,
we also calculated the reionization redshift for different $\zeta$ values. 
In  Fig. \ref{zre2}, 
we plot the contours of $z_{re}$ 
with both $m_{\rm WDM}$ and $\zeta$ varying. 
We see WDM models with particles mass below 0.5~keV have
already been excluded, otherwise the reionization could not be completed before redshift 6 for
reasonable values of $\zeta$. On the other hand, for particles with $m > 8 \keV$,
$\zeta$ should be less than $\approx180$, 
otherwise the completion of reionization would be too early, in conflict with the 
WMAP observations \citep{2011ApJS..192...16L}.

Besides the halo abundance, the distribution of gas within the halo is also affected
by the replacement of CDM with WDM. The collapse of halos in the WDM model 
is generally later than in the CDM model, hence the halo concentration is typically 
smaller \citep{2001ApJ...558..482B,2011PhRvD..84f3507S}, as
the dark matter halos formed later would reach smaller average
densities \citep{1997ApJ...490..493N}.  
In addition to the delayed collapse redshift,
\cite{2011PhRvD..84f3507S} pointed out that due to the relic velocities of WDM particles,
the core of the halo would be smoother in the WDM case. 
These effects could in principle reduce the photon consumption rate of the
minihalos. We have checked this effect by calculation and found that it should only have a very slight
influence on our results. For example, in the $m_{\rm WDM}=10$~keV case, 
for a halo with $10^7~M_\odot$ at $z=10$, the density profile of the halo
 within $4\times10^{-2}~R_{vir}$ would be flattened, 
but outside this core radius the density profile is hardly 
changed. The gas density profile changed even less, since even 
in the CDM model the baryonic gas has pressure and therefore a 
more smooth distribution. The larger minihalos 
contribute more to photon consumption in the WDM model,
because the abundance of smaller ones is reduced significantly. In the end, 
we find that the change in density profile does not significantly affect 
the total recombination rate of minihalos.  

In the present calculation, we have used the bubble model to 
treat the reionization process. 
The bubble model is of course only an approximate model, though it does reproduce 
more elaborated simulations \citep{2007ApJ...654...12Z,2007MNRAS.377.1043M,2011MNRAS.414..727Z}.
Furthermore, the bubble model only provides a 
distribution of bubbles at a given redshift, it does not tell us how each individual 
bubble have grown, so we have made only an {\it static} 
treatment of the recombination process: we calculated the volume of ionized bubble
at a certain redshift, then counted the number of minihalos in this region at that time,
and calculated the number of recombinations for these minihalos until they are all 
photoevaporated. 
This treatment might 
slightly overestimate the number of recombinations, because some of the  
bubbles are formed by the growth and merger of 
smaller bubbles which were formed at earlier time, and 
in these regions the minihalo formation had been suppressed. 
To check whether the neglecting of bubble growth history would alter our 
results, we make a simplifying {\it ansatz} of bubble growth. We assume that for  
a given bubble at redshift $z$, the different parts of its volume were acquired (i.e. 
first ionized) at different redshifts $z'>z$, and the contribution at $z'$ is 
proportional to  
$\frac{df_{coll}}{dz}(z')$. With this assumption on bubble growth 
history, the suppression 
on minihalo formation in previously ionized region can be taken into account, and 
the average number of recombinations within the bubbles is given by
$$\displaystyle \bar{n}_{rec,MH}=\frac{1}{f_{coll}(z)}\int n_{rec,MH(z^\prime)}\frac{df_{coll}(z^\prime)}{dz^\prime}dz^\prime,$$
where $n_{rec,MH}$ on the R.H.S is given by Eq. (\ref{nrecMHs}). 
We then calculate the evolution of $Q_V$ with this new recombination number. 
However, we find that the resulting
difference is very small. This is not surprising, 
since $\frac{df_{coll}}{dz}$ increases rapidly as redshift decreases,
in the calculation of $\bar{n}_{rec,MH}(z)$, 
$n_{rec,MH}(z^\prime\sim z)$ contributes most to the integration.

Finally, \citet{2007Sci...317.1527G} suggested that in the WDM models, unlike the case of the
CDM models, the first stars do not necessarily form in halos, but may instead form
in filaments. They also argued that with this new way of star formation, the first 
stars may have smaller typical mass ($\sim \Msun$), and the global star formation rate 
could be even higher. This also raises the interesting possibility that the reionization 
could occur earlier in the WDM model than in the CDM model.
However, whether this new formation mechanism would indeed work 
as they proposed is still not completely clear, as it is 
very difficult to model the star formation process 
in sufficient resolution at present to 
really check the outcome of the filament star formation process. 
Here, we have maintained the standard view, and assumed that the stars formed only in 
DM halos. We note that if the stars indeed form in the way suggested by 
\citet{2007Sci...317.1527G}, the reionization in the WDM model would occur 
at even higher redshifts, and there is not anything 
incompatible with the recombination effects we discussed.

In summary, we find that in WDM models the completion of reionization is not 
always delayed, in some cases, when the WDM mass is not too low, 
it could even be advanced due to the 
reduction of recombination rates.
This effect is relatively small compared to the uncertainties in observation and theoretical models
at present. We also find that for $\zeta=40$, to be consistent with the observations 
of Gunn-Peterson trough in $z \sim 6$ quasar spectra, 
the mass of dark matter particles should be higher than 
1.3~keV. However, if more ionizing photons were produced and 
escaped into the IGM, the dark matter could be warmer, i.e have smaller mass. 
However, for dark matter particles 
that are less than 0.5~keV, $\zeta \gtrsim 500$ is needed. This would only be possible if 
the reionization photons are contributed primarily by metal-free or massive extremely metal-poor 
stars.

\section*{ACKNOWLEDGMENTS}
We thank Yidong Xu and Liang Gao for helpful discussions. 
This work is supported by the 973 project under grant 2007CB815401, 
NSFC grant No.11073024, the John Templeton Foundation,
and the CAS knowledge innovation program.


\end{document}